\documentclass[a4paper]{PoS}
\usepackage[utf8]{inputenc}
\usepackage{multirow}
\usepackage{booktabs}
\usepackage{amsmath}
\DeclareMathOperator {\tr}{tr}

\newcommand{\umux}{\ensuremath{U_\mu(x)}}

\newcommand{\vmux}{\ensuremath{V_\mu(x)}}

\newcommand{\qmux}{\ensuremath{q_\mu(x)}}

\newcommand{\Dms}{\ensuremath{D_\mu^*}}

\newcommand{\chq}{\ensuremath{\cosh\{a g_0 q_\mu(x)\}}}
\newcommand{\shq}{\ensuremath{\sinh\{a g_0 q_\mu(x)\}}}

\title{Numerical Stochastic Perturbation Theory in the Schrödinger Functional}

\ShortTitle{NSPT in the SF}

\author{Michele Brambilla$^a$, Mattia Dalla Brida$^b$, Francesco Di Renzo$^a$,
  \vfill\speaker{Dirk Hesse}$^{,a,1}$, and Stefan Sint$^{b,c}$\vspace{.5em}\\
  $^a$ Università degli studi di Parma, Viale G.P. Usberti 7/a,
  43100 Parma, Italy and INFN\\
  $^b$ School of Mathematics, Trinity College Dublin, Dublin 2,
  Ireland\\
  $^c$NIC@DESY Zeuthen, Platanenallee 6, 15738 Zeuthen, Germany\\
  $^1$ E-Mail: dirk.hesse@fis.unipr.it
}

\abstract{The Schrödinger functional (SF) is a powerful and widely
  used tool for the treatment of a variety of problems in
  renormalization and related areas. Albeit offering many conceptual
  advantages, one major downside of the SF scheme is the fact that
  perturbative calculations quickly become cumbersome with the
  inclusion of higher orders in the gauge coupling and hence the use
  of an automated perturbation theory framework is desirable. We
  present the implementation of the SF in numerical stochastic
  perturbation theory (NSPT) and compare first results for the running
  coupling at two loops in pure $SU(3)$ Yang-Mills theory with the
  literature.}

\FullConference{31st International Symposium on Lattice Field Theory - LATTICE 2013\\
		July 29 - August 3, 2013\\
		Mainz, Germany}

\begin{document}

\section{Numerical Stochastic Perturbation Theory and the Schrödinger
  Functional}

We consider the Schödinger functional as defined in
\cite{Luscher:1992an}. Specifically, the $SU(3)$ gauge field variables
$U_\mu(x)$ are subject to periodic boundary conditions in the spatial
directions and Dirichlet ones in the temporal directions,
\begin{equation}
  \label{eq:1}
  U_k(x) |_{x_0 = 0} = e^{a C_k}\,,\quad U_k(x)|_{x_0 = T} = e^{a C_k'}\,.
\end{equation}
The presence of Dirichlet boundary conditions induces a background
field as was discussed at length in \cite{Luscher:1992an}, which can
be exploited for a number of interesting applications, most
prominently for the determination of the running gauge coupling as we
will shortly review. The most common choices of the matrices $C_k,
C_k'$ are either zero, in which case we will speak of a
\emph{trivial} background field, and constant diagonal matrices as
specified in \cite{Luscher:1993gh}, which gives raise to an
\emph{Abelian} background field,
\begin{equation}
  \label{eq:2}
  V_\mu(x) = e^{aB_\mu(x)}\,,\quad B_0 = 0\,,\quad B_k = \left[x_0
    C_k' + (T - x_0) C_k\right]/L\,.
\end{equation}
Close to the classical minimum, the gauge fields may then be written
as
\begin{equation}
  \label{eq:9}
  \umux = e^{a\, g_0\,\qmux} \vmux\,.
\end{equation}
The gauge action is given by
\begin{equation}
  \label{eq:4}
  S[U] = \frac 1 {g_0^2} \sum_p w(p) \tr \left\{ 1 - U(p) \right\}\,,
\end{equation}
where we sum over all oriented plaquettes $p$, $U(p)$ is the closed
parallel transporter around $p$, and the weight factor is unity
everywhere on the lattice except for the plaquettes that contain a
frozen spatial link at the boundary (\ref{eq:1}) and a temporal link,
for which we set
\begin{equation}
  \label{eq:5}
  w(p)|_{\rm boundary} = c_t = 1 + g_0^2 c_t^{(1)} + g_0^4 c_t^{(2)} + \ldots\,,
\end{equation}
with $c_t^{(1)} = -0.08896(23)$ as quoted in \cite{Bode:1998hd}. This
ensures cancellation of $O(a)$ cut-off effects introduced by the
boundary.

Numerical stochastic perturbation theory is reviewed in detail in
\cite{DiRenzo:2004ge}. It is a numerical implementation of stochastic
perturbation theory, which solves the equations of stochastic
quantization order by order in the coupling constant of the
theory. Specifically, one introduces a new degree of freedom, the
\emph{stochastic time} $t$ and considers solutions of the Langevin
equation,
\begin{equation}
  \label{eq:3}
  \partial_t U^\eta_\mu(x; t) = \left\{-i\,\nabla_{x\mu} S[U] -
    i\eta_{x\mu}(t)\right\} U^\eta_\mu(x; t)\,,
\end{equation}
where the gauge field now acquires a formal dependence on the choice
of the Gaussian noise variables $\eta(t)=T^a\eta^a$, which in turn
satisfy
\begin{align}
\big\langle\eta^a_{x\mu}(t)\big\rangle_\eta & = 0\,, &
\big\langle\eta^a_{x\mu}(t)\eta^b_{y\nu}(t')\big\rangle_\eta & = 
2\,\delta^{ab}\delta_{xy}\delta_{\mu\nu}\delta(t-t')\,.
\end{align}
The main assertion of stochastic quantization
(\cite{Damgaard:1987rr,Parisi:1980ys}) is that the noise average of
any (in the case at hand, gauge-invariant) observable $O(U)$
converges, in the limit of large stochastic time, to the usual path
integral average
\begin{equation}
  \label{eq:6}
  \big\langle O(U_\mu^{\eta}) \big\rangle_\eta \xrightarrow{t \to
    \infty} \big\langle O \big\rangle\,.
\end{equation}
In particular in the case of non-Abelian gauge theories the proof of
(1.8) is non-trivial, but it is rigorous in perturbation theory
\cite{Floratos:1982xj}. Hence the solution to the Langevin equation
may be consistently expanded in the coupling constant
\begin{equation}
  \label{eq:7}
  U_\mu(x) = V_\mu(x) + \sum_{i=1}^N g_0^i\,U_\mu^{(i)}(x) + O\left(g_0^{N+1}\right)\,.
\end{equation}
NSPT amounts to storing the perturbative expansion of the gauge field
on a computer and numerically solving the order-by-order version of
the Langevin equation. This is done using e.g. an Euler or Runge-Kutta
integrator. During integration, all multiplications of gauge fields
(\ref{eq:7}) have to be performed order-by-order, which may be
conveniently implemented using any modern object-oriented programming
language. We choose to employ the second-order Runge-Kutta method
described in \cite{Davies:1985ad}, introducing discretization errors
proportional to the squared integration step-size $\tau^2$. These
cut-off effects have to be extrapolated to zero in the final
analysis. The form of the expansion (\ref{eq:7}) makes it clear that
trivial as well as Abelian background fields may be accommodated. A
more detailed description of our implementation may be found in
\cite{PoS:Michele}.

\subsection{Stochastic Gauge Fixing}

\begin{figure}[ht]
  \centering
  \includegraphics{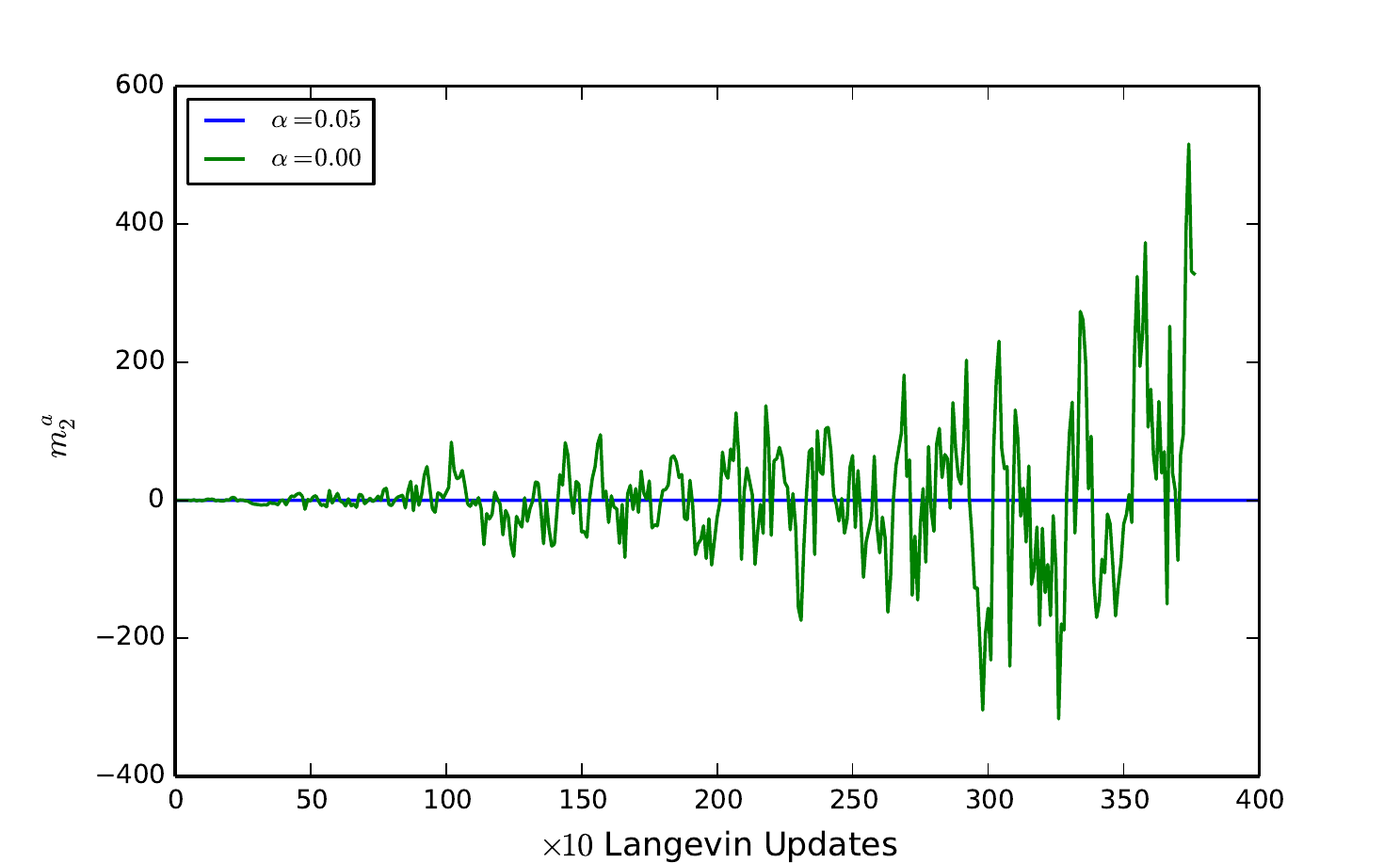}
  \caption{The effect of stochastic gauge fixing.}
  \label{fig:noise}
\end{figure}
Zwanziger described in \cite{Zwanziger:1981kg} a method to perform
gauge fixing in a theory that has been quantized stochastically. As
explained in \cite{DiRenzo:2004ge}, a similar approach, which we want
to call \emph{gauge suppression} here, is crucial to extract results
form NSPT. The noise term in the Langevin equation (\ref{eq:3}) drives
all modes of the gauge field, while only the transversal ones are then
damped by the term proportional to the group derivative of the gauge
action. The longitudinal modes, even though they do not contribute to
expectation values of gauge-invariant quantities, need to be damped as
well if one hopes to extract numerical results, which are otherwise
drowned out in noise. To construct a gauge transformation which will
damp the unwanted modes with SF boundary conditions present, we
proceed in the usual fashion (see e.g.\ \cite{Giusti:2001xf} for a
review) by finding a suitable functional $W[U]$ whose extremal value
is characterized by the gauge fixing condition (as specified in
\cite{Luscher:1992an}) being fulfilled. Our choice is given by
\begin{equation}
  \label{eq:8}
    W[U] = -\,a^2  \sum_{(x,\mu)\in \Lambda} \chq \,,
\end{equation}
where $\Lambda$ is the set of all indices belonging to
\emph{dynamical} links. The functional (\ref{eq:8}) is driven to its
stationary point by applying gauge fixing transformations $\Omega(x) =
e^{-i\, \epsilon\, \omega(x)}$ with, in the case of an \emph{Abelian}
background field,
\begin{equation}
  \label{eq:11}
  \omega(x) = \begin{cases}
    - \alpha \sum_{\mu} \left. \Dms 
      \shq \right|_{\rm traceless}\,, &{\rm if~}\, 0 < x_0 < T  \,,\\
    - \alpha \left( \frac a L \right)^3 
    \sum_{x, x_0 = 0} \left. \sinh\{a g_0 q_0(x) \}_{jj} 
    \right|_{\rm traceless}\,, & {\rm if~}\, x_0 = 0\,\\
    0 &{\rm else}\,,
  \end{cases}
\end{equation}
with $0<\alpha < 1$, $j$ being a color index (i.e.\ we refer to the
diagonal, traceless part in the second line of \eqref{eq:11}), and the
covariant derivative as specified in \cite{Luscher:1992an}. In the
case of a trivial background field, one has to consider the full color
matrix at $x_0 = 0$. To demonstrate the high sensitivity of NSPT
simulations on the gauge suppression being in place, we performed two
runs, starting from the same thermalized configuration on a $L/a = 4$
lattice with integration step size $\tau = 0.02$, one with gauge
suppression in place and another one with $\alpha = 0$. The effect is
rather dramatic, even tough the mean value remains the same, the
errors explode due to ever-growing fluctuations, as seen in figure
\ref{fig:noise}. The observable considered there will be introduced in
the following section.

\section{The Gauge Coupling}

\begin{figure}[ht]
  \centering
  \includegraphics{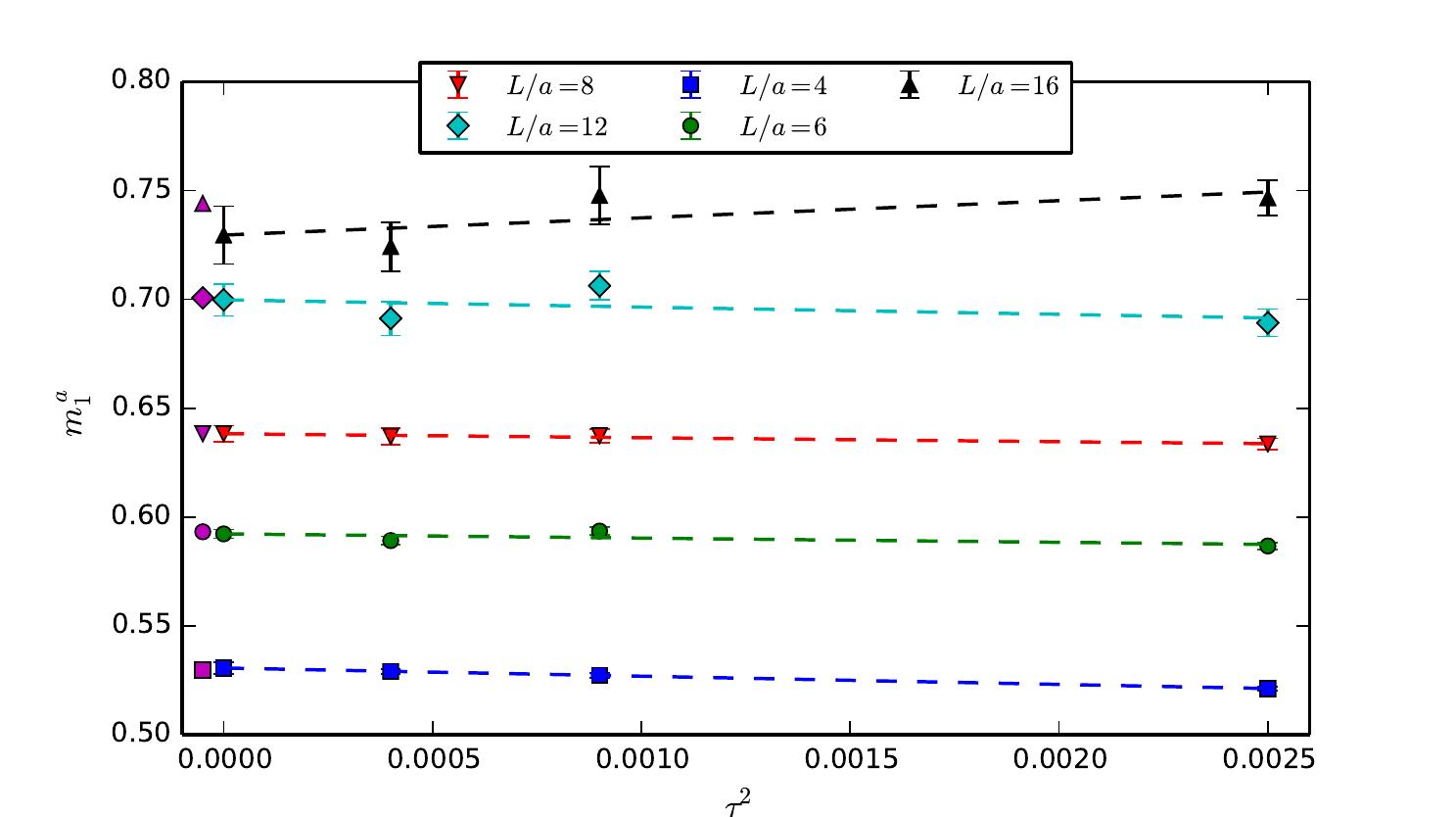}
  \includegraphics{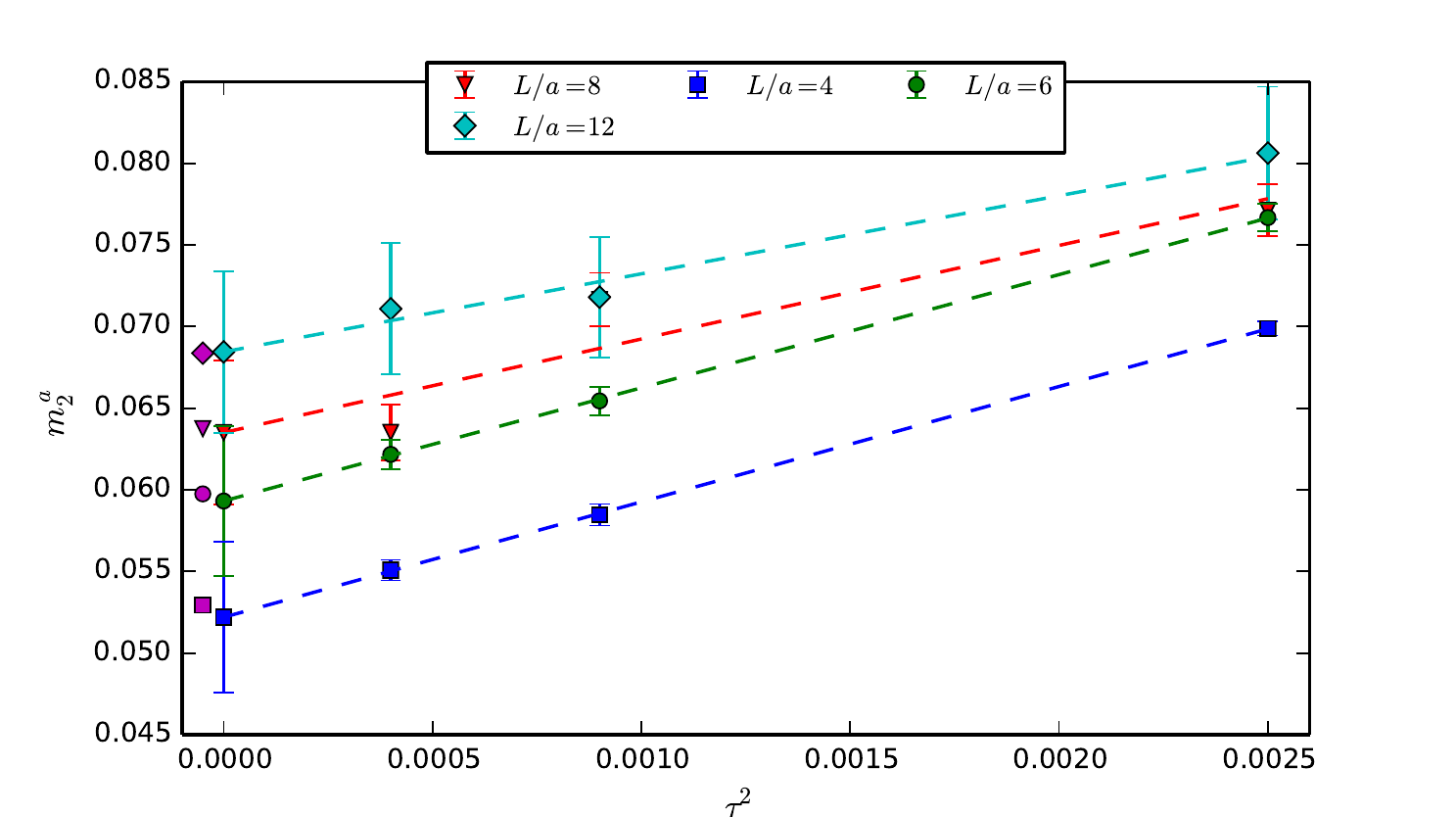}
  \caption{Results for $m_1^a$ and $m_2^a$, the magenta points represent the known
    results.}
  \label{fig:m1ab}
\end{figure}

As a suitable test case to reliably check the correctness of our
implementation, we chose to re-calculate the two-loop results of the
Schrödinger functional coupling presented in \cite{Bode:1998hd}. The
boundary values for the gauge fields chosen in \cite{Luscher:1993gh}
depend on a parameter $\eta$, which may be used to define a running
coupling $\overline g$ using the free energy
\begin{equation}
  \label{eq:12}
  e^{-\Gamma} = \int D{U}e^{-S[U]}\,,
\end{equation}
by setting
\begin{equation}
  \label{eq:13}
  \overline g^2 = k / \partial_\eta \Gamma = g_0^2 + m_1\,g_0^4 +
  m_2\,g_0^6 + \ldots\,,
\end{equation}
where the constant $k$ ensures correct normalization. As in
\cite{Bode:1998hd}, we will adapt a notation to single out the
contributions coming from the weight factor (\ref{eq:5}), writing
\begin{align}
  \label{eq:14}
  m_1 =\; &m_1^a + c_t^{(1)}\,m_1^b\,,\\
  \label{eq:15}
  m_2 - m_1^2 =\;& m_2^a + c_t^{(1)}\,m_2^b +
  \left(c_t^{(1)}\right)^2\,m_2^c + c_t^{(2)}\,m_2^d\,.
\end{align}
All individual parts of (\ref{eq:14}) and (\ref{eq:15}) were
calculated in \cite{Bode:1998hd} for a range of lattice resolutions
$L/a$. First, we neglect the effect of boundary improvement and set
$c_t = 1$, focusing on reproducing $m_1^a$ and $m_2^a$. We performed a
number of exploratory simulations on rather small lattices with $L/a$
ranging from four to 16. We did not find it worthwhile to push the
lattice sizes further since our main aim is to prove the correctness
of our implementation of the Schrödinger functional in NSPT. To do so,
we proceed as follows. For each chosen lattice size $L/a$, we
performed three simulations varying the integration step size $\tau
\in \{0.02, 0.03, 0.05\}$. We estimated the mean value and variance by
taking into account the integrated autocorrelation time, estimated as
recommended in \cite{Wolff:2003sm}. We then extrapolated $\tau \to 0$
using both a linear fit to all data points and a constant fit to the
data omitting $\tau = 0.05$. The difference in the fits is then added
to the statistical error as an estimate for systematic effects in the
extrapolation, giving the NSPT estimate $m_i^{a,\rm NSPT}$ and error
$\delta m_i^{a,\rm NSPT}$. The results for $m_1^a$ and $m_2^a$ may be
found in figure \ref{fig:m1ab}. We define the \emph{precision} of our
estimate by comparing with the known results,
\begin{equation}
  \label{eq:16}
  \Delta m_i^a = \left|m_i^a - m_i^{a,\rm NSPT}\right|\,.
\end{equation}
\begin{table}[ht]
  \centering\scriptsize
  \begin{tabular}{cc|cc|cc|cc|cc|cc}
\toprule
\midrule\multicolumn{2}{c}{$L/a$} & \multicolumn{2}{|c}{4} & \multicolumn{2}{|c}{6} & \multicolumn{2}{|c}{8} & \multicolumn{2}{|c}{12} & \multicolumn{2}{|c}{16}\\
\midrule \multicolumn{2}{c|}{$\tau$} & 0.02 & 0.03 & 0.02 & 0.03 & 0.02 & 0.03 & 0.02 & 0.03 & 0.02 & 0.03\\
\midrule\multirow{3}{*}{$g_0^2$} & $N_{\rm eff}$ & 11466 & 16760 & 5851 & 8242 & 2256 & 3804 & 864 & 1363 & 478 & 353\\
&$\tau_{\rm int}$ & 12.6(4) & 8.6(2) & 41(2) & 29(1) & 66(4) & 39(2) & 83(8) & 53(4) & 92.9(1) & 62(8)\\
\cmidrule{2-12}&$\Delta m_1^a$ & \multicolumn{2}{|c}{0.0009(27)} & \multicolumn{2}{|c}{0.0010(21)} & \multicolumn{2}{|c}{0.0000(37)} & \multicolumn{2}{|c}{0.0009(74)} & \multicolumn{2}{|c}{0.014(13)}\\
\midrule\multirow{3}{*}{$g_0^4$} & $N_{\rm eff}$ & 12489 & 17334 & 6105 & 8545 & 2544 & 3633 & 825 & 1181 & 327 & 221\\
&$\tau_{\rm int}$ & 11.5(3) & 8.3(2) & 39(2) & 27.9(10) & 59(3) & 41(2) & 87(8) & 61(5) & 135.9(2) & 99.2(2)\\
\cmidrule{2-12}&$\Delta m_2^a$ & \multicolumn{2}{|c}{0.0007(46)} & \multicolumn{2}{|c}{0.0004(46)} & \multicolumn{2}{|c}{0.0002(44)} & \multicolumn{2}{|c}{0.0001(49)} & \multicolumn{2}{|c}{0.0032(88)}\\
\bottomrule
\end{tabular}

  \caption{Details of simulations performed.}
  \label{tab:details}
\end{table}
The details of our simulations are summarized in table
\ref{tab:details}, where we omitted the data points at the biggest
integration step size for space reasons (the performance is anyway
limited by the simulation with smallest step size due to larger
autocorrelation times). In addition to the integrated autocorrelation
time in units of Langevin integration steps and precision of our
estimation, we state the effective number of measurements $N_{\rm eff}
= N_{\rm updates} / (2\tau_{\rm int})$. The total run-time for $L/a =
12$, leading to a result of $m_2^a = 0.0684(49)$ was 210h on a single
node of the Lonsdale cluster at Trinity College (AMP Opteron, 8
cores/node).

\subsection{Boundary Improvement}
\label{sec:boundary-improvement}

\begin{table}[ht]
  \centering\scriptsize
  \begin{tabular}{c c | c c c | c c c | c c c}
\toprule
\multicolumn{2}{c}{$L/a$} & \multicolumn{3}{|c}{4} & \multicolumn{3}{|c}{6} & \multicolumn{3}{|c}{8}\\
\midrule \multicolumn{2}{c|}{$\tau$} & 0.02 & 0.032 & 0.04 & 0.02 & 0.032 & 0.04 & 0.02 & 0.032 & 0.04\\
\midrule\multirow{3}{*}{$g_0^2$} & $N_{\rm eff}$ & 6418 & 6357 & 6225 & 2775 & 3069 & 2776 & 1286 & 1266 & 1193\\
&$\tau_{\rm int}$ & 13.3(5) & 7.8(3) & 6.7(3) & 21(1) & 13.0(7) & 10.7(6) & 28(2) & 16(1) & 15(1)\\
\cmidrule{2-11}&$\Delta m_1$ & \multicolumn{3}{|c}{0.0026(64)} & \multicolumn{3}{|c}{0.0007(27)} & \multicolumn{3}{|c}{0.0030(39)}\\
\bottomrule
\end{tabular}

  \caption{Details of simulations performed including $c_t^{(1)}$.}
  \label{tab:ct1-details}
\end{table}

%

For many applications, boundary improvement is a necessity. Hence we
performed a few additional simulations, now including $c_t^{(1)}$ in
the gauge update step. Note that up to $O(g_0^6)$ this is sufficient
to capture all non-tree-level dynamic, as the contribution involving
$c_t^{(2)}$ multiplies a tree-level counter-term. The results are
presented in 
table \ref{tab:ct1-details} and are found to be in complete agreement
with what was found in \cite{Bode:1998hd}. We did not attempt to
extract $m_2$ due to limited statistics.

\section{Conclusions}

The availability of the Schrödinger functional in NSPT opens the door
to a variety of interesting applications, in particular connected to
the Wilson flow method \cite{Luscher:2010iy}, as will be further
discussed in \cite{PoS:Mattia}. We are confident in the correctness
and performance of our code, using trivial as well as Abelian
background fields. The inclusion of dynamical fermions is well on the
way.

\subsection*{Acknowledgements}

We would like to thank the teams of the computing center at Trinity
College (TCHPC) and the Irish Centre for High End Computing (ICHEC)
for their support in performing the numerical simulations. S.\ Sint
acknowledges support by SFI under grant 11/RFP/PHY3218.  M.\ Dalla
Brida is supported by the Irish Research Council. This work received
funding from the Research Executive Agency (REA) of the European Union
under Grant Agreement number PITN-GA-2009-238353 (ITN STRONGnet) and
in parts by INFN under i.s.\ MI11 (now QCDLAT).

\bibliographystyle{unsrt}
\bibliography{lat13}

\end{document}